\documentclass{article}
\usepackage[utf8]{inputenc}
\usepackage[sort,compress,numbers]{natbib}
\usepackage{graphicx,xcolor}
\usepackage{subfigure}
\usepackage{amsmath}
\newcommand{\be}{\begin{equation}
\newcommand{\ee}{\end{equation}}}
\newcommand{\bea}{\begin{eqnarray}}
\newcommand{\eea}{\end{eqnarray}}
\newcommand{\nn}{\nonumber}
 
\begin{document}

%
%
\title{The effect of deformation of special relativity by conformable derivative}
\author{Ahmed Al-Jamel$^1$, Mohamed.Al-Masaeed$^1$, Eqab.M.Rabei$^1$, Dumitru Baleanu$^2$ \\
$^1$ Physics Department, Faculty of Science,\\ Al al-Bayt University, P.O. Box 130040, Mafraq 25113, Jordan
\\
$^2$Department of Mathematics, Cankaya University, Ankara, Turkey,\\
\\aaljamel@aabu.edu.jo, aaljamel@gmail.com\\moh.almssaeed@gmail.com\\eqabrabei@gmail.com\\dumitru@cankaya.edu.tr
}

\markboth{ Ahmed Al-Jamel}{}

%
%

\maketitle


\begin{abstract}
In the article, the deformation of special relativity within the frame of conformable derivative is formulated. Within this context, the two postulates of the theory were re-stated. And, the addition of velocity laws were derived and used to verify the constancy of the speed of light.  
The invariance principle of the laws of physics is demonstrated for some typical illustrative examples, namely, the conformable wave equation, the conformable Schrodinger equation, and the conformable Gordon-Klein equation. The current formalism may be applicable when using special relativity in a nonlinear or dispersive medium.
\\

\textit{Keywords:} conformable derivative, fractional calculus, special relativity.
\end{abstract}

\section{Introduction}
The Einstein's special relativity plays a corner stone in modern physics. As stated in 1905 by Einstein, it is based on two postulates. The first postulate is about the constancy of the speed of light: the speed of light $c$ is the same in all inertial frames of references. The second postulate is about the invariance form of the laws of physics under Lorentz transformations. As a consequence of this, any theory of space and time should be compatible with the theory of special relativity. There are some other aspects that were studied after the emergence of theory of relativity. In \cite{michels1941special}, the Lorentz transformations were re-stated for an observer in a refracting but non-dispersive medium was proposed, and some physical consequences were discussed. In \cite{crenshaw2019reconciliation}, Laue and Rosen theories of dielectric special relativity were derived, and argued that both are true but with different range of applicability. In \cite{mashhoon2012nonlocal}, the non-local special relativity is introduced to overcome the difficulties accompanied the non-local electrodynamics problems. 
\\

In the last two decades, the fractional calculus approach to model or resolve various physical problems has attracted many researchers. There are a number of definitions or senses for fractional calculus such as Riemann-Liouville, Caputo, Riesz and Weyl \cite{podlubny1998fractional,oldham1974fractional,kilbas2006theory,miller1993introduction}. In 2014, Khalil et.al \cite{khalil2014new} proposed a modern definition for fractional calculus called conformable fractional derivative (CFD). For a given function
 $f(t)\in [0,\infty) \to {R}$, the conformable derivative of $f(t)$ of order $\alpha$, denoted as $D_{t}^\alpha f(t)$ is defined as \cite{khalil2014new}:
\be
\label{conformable}
D_{t}^\alpha f(t)=\lim_{\epsilon \to 0}\frac{f(t+\epsilon t^{1-\alpha})-f(t)}{\epsilon}.
\ee
This definition is simple in the sense that it satisfies the general properties of the ordinary derivative including the Leibniz and chain rules.
In \cite{abdeljawad2015conformable,atangana2015new}, this CFD is re-investigated and new properties similar to these in traditional calculus were derived and discussed. The CFD has been used to study various physical problems with possible nonlinear or diffusive nature. In \cite{al2019search}, the mass spectroscopy of heavy mesons were investigated within the frame of conformable derivative searching for any ordering effect in their spectra that varies with the fractional order. In \cite{tarasov2010fractional}, the fractional dynamics of relativistic particles was studied, and it was found that fractional dynamics of such particles are described as non-Hamiltonian and dissipative. Possibility of being Hamiltonian system under some conditions was also presented. In \cite{chung2021new},  a new conformable fractional mechanics using the fractional addition was proposed and new definitions for the fractional velocity fractional acceleration are given. In \cite{chung2020effect}, deformation of quantum mechanics due to the inclusion of conformable fractional derivative is presented and investigated with some physical illustrative examples. In \cite{AlMasaeedRabeiAlJamelBaleanu+2021+395+401},  the Hamiltonian for the conformable harmonic oscillator is constructed using fractional operators termed $\alpha$-creation and $\alpha$-annihilation operators. The exact voyaging (2 + 1) dimensional Heisenberg ferromagnetic spin chain solutions with conformable fractional derivatives were investigated in  \cite{akhtar2021variety}. Recently, Pawar et.al \cite{pawar2018approach} introduced Riemannian Geometry through using the conformable fractional derivative in Christoffel index symbols of the first and second kind. Later, in \cite{abdelhakimFlawConformableCalculus2019}, pointed out the conformable derivative is not fractional but it is an operator. Thus in the present paper we call it conformable derivative.
\\
The purpose of this paper is to investigate the deformation of the theory of special relativity within the frame of conformable fractional derivative. This means that, we will construct a new set of Lorentz transformations, re-state the postulates of special relativity, and then verify the validity of the invariance principle to various laws or equations of physics.

\section{Theory}
Deformation of Lorentz transformations using conformable derivative is reported in \cite{chung2020effect}.

\textbf{Definition} The $\alpha-$ Lorentz transformations between two inertial frames $S$ and $S'$ are defined as \cite{chung2020effect}:
\be
\label{x'}
x'^\alpha =\Gamma_\alpha(x^\alpha-v_\alpha t^\alpha).
\ee
\be
\label{t'}
t'^\alpha =\Gamma_\alpha(t^\alpha-\frac{v_\alpha}{c^{2\alpha}} x^\alpha).
\ee
\be
\label{y'}
y'^\alpha = y^\alpha.
\ee
\be
\label{z'}
z'^\alpha = z^\alpha,
\ee
where $\Gamma_\alpha=\frac{1}{\sqrt{1-\frac{v_\alpha^2}{c^{2\alpha}}}}$ is the $\alpha-$ deformed Einstein factor and $v_\alpha$  is the $\alpha-$relative velocity between the two frames.

We now state the two postulates of conformable special relativity as follows.
\begin{itemize}
	\item Postulate 1(Constancy of the speed of light): The speed of light is the same for all $\alpha-$inertial frames of references.
	\item Postulate 2(Invariance Principle): The laws of physics are invariant under $\alpha-$Lorentz transformations.
\end{itemize}
The following subsections purpose is to clarify theses two postulates.
\subsection{The $\alpha-$velocity addition law}
Following \cite{chung2021new}, we define the $\alpha-$velocity of an event with respect to the $S$ and $S'$ frames as
\bea
\label{v_alpha}
u_\alpha \equiv D_{t}^\alpha x^\alpha &=& (\frac{t}{x})^{1-\alpha} \frac{dx}{dt} \\
\label{v'_alpha}
u'_\alpha \equiv D_{t'}^\alpha x'^\alpha &=& (\frac{t'}{x'})^{1-\alpha} \frac{dx'}{dt'},
\eea
respectively. To calculate the velocity using eq.\eqref{x'} and \eqref{t'}, we have
\bea
\nn
\frac{dx'^\alpha}{dt'^\alpha} =\frac{\Gamma_\alpha(dx^\alpha-v_\alpha dt^\alpha)}{\Gamma_\alpha(dt^\alpha-\frac{v_\alpha}{c^{2\alpha}} dx^\alpha)}=\frac{(\frac{dx^\alpha}{dt^\alpha}-v_\alpha )}{(1-\frac{v_\alpha}{c^{2\alpha}} \frac{dx^\alpha)}{dt^\alpha}}.
\eea
By interpreting $\frac{dx'^\alpha}{dt'^\alpha}=u'_\alpha$ and $\frac{dx^\alpha}{dt^\alpha}=u_\alpha$, we thus obtain
\bea
u'_\alpha =\frac{(u_\alpha -v_\alpha )}{(1-\frac{v_\alpha}{c^{2\alpha}}u_\alpha ) }.
\eea

In case $u_x = c$, we have 
\bea
(\frac{x'}{t'})^{\alpha-1} u'_x =\frac{((\frac{x}{t})^{\alpha-1} u_x -v_\alpha )}{(1-\frac{v_\alpha}{c^{2\alpha}}(\frac{x}{t})^{\alpha-1} u_x ) } = \frac{((\frac{x}{t})^{\alpha-1} c -v_\alpha )}{(1-\frac{v_\alpha}{c^{2\alpha}}(\frac{x}{t})^{\alpha-1} c ) },
\eea
where we have made use of eq.\eqref{v_alpha} and \eqref{v'_alpha}. With the realization $\frac{x}{t} = c$ and $\frac{x'}{t'}= c $, we have
\bea
c^{\alpha-1} u'_x = \frac{(c^{\alpha-1} c -v_\alpha )}{(1-\frac{v_\alpha}{c^{2\alpha}}c^{\alpha-1} c ) } = \frac{(c^{\alpha} -v_\alpha )}{(1-\frac{v_\alpha}{c^{\alpha}}  ) }= c^{\alpha} \frac{(c^{\alpha} -v_\alpha )}{(c^{\alpha} -v_\alpha )},
\eea
from which we obtain
\bea
c^{\alpha-1} u'_x = c^{\alpha} \to u'_x = c^{1- \alpha}c^{\alpha} = c,
\eea
or 
\bea
 u'_x = c.
\eea
This verifies that the $\alpha-$Lorentz transformations proposed in Eq. (\ref{x'}-\ref{z'}) leads to the constancy of the speed of light.
\subsection{Conformable wave equation} 
Here, we test covariance of the wave equation under the $\alpha-$ Lorentz transformation. The $\alpha-$ wave equation in $ 1+1$ dimension  \cite{chung2020effect}, we have 
\be
\label{wave eq 1d}
D^\alpha_x D^\alpha_x \Psi- \frac{1}{c^{2\alpha}} D^\alpha_t D^\alpha_t \Psi=0.
\ee
Using the $\alpha-$Laplacian \cite{mhailan2020fractional}, we have then 
\be
\label{wave  eq 3d}
\nabla^{2\alpha} \Psi - \frac{1}{c^{2\alpha}} D^\alpha_t D^\alpha_t \Psi =0,
\ee
where $\nabla^{2\alpha} = D^\alpha_x D^\alpha_x + D^\alpha_y D^\alpha_y + D^\alpha_z D^\alpha_z$. Using of the chain rule \cite{atangana2015new}
\bea
\nn
D^\alpha_x \Psi = x'^{\alpha-1} D^\alpha_x x' D^\alpha_{x'} \Psi + t'^{\alpha-1} D^\alpha_x t' D^\alpha_{t'} \Psi,
\eea
and then using the $\alpha-$Lorentz transformations eq.\eqref{x'} and \eqref{t'}, $x' =\Gamma_\alpha^{\frac{1}{\alpha}} (x^\alpha-v_\alpha t^\alpha)^{\frac{1}{\alpha}}, t' =\Gamma_\alpha^{\frac{1}{\alpha}} (t^\alpha-\frac{v_\alpha}{c^{2\alpha}} x^\alpha)^{\frac{1}{\alpha}}$,  we have
\bea
\nn
D^\alpha_x \Psi &=& (\Gamma_\alpha^{\frac{1}{\alpha}} (x^\alpha-v_\alpha t^\alpha)^{\frac{1}{\alpha}})^{\alpha-1} x^{1-\alpha} \frac{d}{dx}\Gamma_\alpha^{\frac{1}{\alpha}} (x^\alpha-v_\alpha t^\alpha)^{\frac{1}{\alpha}} D^\alpha_{x'} \Psi \\\nn
&+& (\Gamma_\alpha^{\frac{1}{\alpha}} (t^\alpha-\frac{v_\alpha}{c^{2\alpha}} x^\alpha)^{\frac{1}{\alpha}})^{\alpha-1} x^{1-\alpha} \frac{d}{dx}\Gamma_\alpha^{\frac{1}{\alpha}} (t^\alpha-\frac{v_\alpha}{c^{2\alpha}} x^\alpha)^{\frac{1}{\alpha}} D^\alpha_{t'} \Psi, \\\nn
&=& \Gamma_\alpha^{1-\frac{1}{\alpha}} (x^\alpha-v_\alpha t^\alpha)^{1-\frac{1}{\alpha}} x^{1-\alpha} \Gamma_\alpha^{\frac{1}{\alpha}}  \frac{1}{\alpha}(x^\alpha-v_\alpha t^\alpha)^{\frac{1}{\alpha}-1} \alpha x^{\alpha-1} D^\alpha_{x'} \Psi\\\nn
&-& \Gamma_\alpha^{1-\frac{1}{\alpha}} (t^\alpha-\frac{v_\alpha}{c^{2\alpha}} x^\alpha)^{1-\frac{1}{\alpha}} x^{1-\alpha} \Gamma_\alpha^{\frac{1}{\alpha}} \frac{1}{\alpha}(t^\alpha-\frac{v_\alpha}{c^{2\alpha}} x^\alpha)^{\frac{1}{\alpha}-1}\frac{v_\alpha}{c^{2\alpha}} \alpha x^{\alpha-1} D^\alpha_{t'} \Psi\\\label{first x}
&=&\Gamma_\alpha  D^\alpha_{x'} \Psi-\Gamma_\alpha \frac{v_\alpha}{c^{2\alpha}}D^\alpha_{t'} \Psi.
\eea
Operating again on $D^\alpha_x \Psi$ by $D^\alpha_x$, we have
\bea
\nn
D^\alpha_x D^\alpha_x \Psi &=&(\Gamma_\alpha  D^\alpha_{x'}-\Gamma_\alpha \frac{v_\alpha}{c^{2\alpha}}D^\alpha_{t'})  (\Gamma_\alpha  D^\alpha_{x'} \Psi-\Gamma_\alpha \frac{v_\alpha}{c^{2\alpha}}D^\alpha_{t'} \Psi), \\\label{2nd x}
&=& \Gamma_\alpha^2  D^\alpha_{x'}  D^\alpha_{x'} \Psi - 2 \Gamma_\alpha^2 \frac{v_\alpha}{c^{2\alpha}}  D^\alpha_{x'} D^\alpha_{t'} \Psi + \Gamma_\alpha^2 \frac{v_\alpha^2}{c^{4\alpha}} D^\alpha_{t'} D^\alpha_{t'}\Psi.
\eea
From eqs. \eqref{y'} and \eqref{z'}, it is clear that 
\bea
\nn
y'^\alpha &=& y^\alpha \to \alpha y'^{\alpha-1} dy'= \alpha y^{\alpha-1} dy \to  y'^{1-\alpha} \frac{d}{dy'}= y^{1-\alpha} \frac{d}{dy},
\eea
and thus 
\be
\label{first y'}
D_{y'}^\alpha = D_{y}^\alpha.
\ee
So, 
\be
\label{2nd y'}
D_{y'}^\alpha D_{y'}^\alpha = D_{y}^\alpha D_{y}^\alpha .
\ee
Similarly, 
\be
\label{first z'}
D_{z'}^\alpha = D_{z}^\alpha,
\ee
and  
\be
\label{2nd z'}
D_{z'}^\alpha D_{z'}^\alpha = D_{z}^\alpha D_{z}^\alpha.
\ee
Similar procedure for the $t$ dependence of eq.(\ref{wave  eq 3d}). We implement the chain rule \cite{atangana2015new}: 
\bea
\nn
D^\alpha_t \Psi &=& x'^{\alpha-1} D^\alpha_t x' D^\alpha_{x'} \Psi + t'^{\alpha-1} D^\alpha_t t' D^\alpha_{t'} \Psi,\\\nn
&=& (\Gamma_\alpha^{\frac{1}{\alpha}} (x^\alpha-v_\alpha t^\alpha)^{\frac{1}{\alpha}})^{\alpha-1} t^{1-\alpha} \frac{d}{dt}\Gamma_\alpha^{\frac{1}{\alpha}} (x^\alpha-v_\alpha t^\alpha)^{\frac{1}{\alpha}} D^\alpha_{x'} \Psi \\\nn
&+& (\Gamma_\alpha^{\frac{1}{\alpha}} (t^\alpha-\frac{v_\alpha}{c^{2\alpha}} x^\alpha)^{\frac{1}{\alpha}})^{\alpha-1} t^{1-\alpha} \frac{d}{dt}\Gamma_\alpha^{\frac{1}{\alpha}} (t^\alpha-\frac{v_\alpha}{c^{2\alpha}} x^\alpha)^{\frac{1}{\alpha}} D^\alpha_{t'} \Psi, \\\nn
&=& -\Gamma_\alpha^{1-\frac{1}{\alpha}} (x^\alpha-v_\alpha t^\alpha)^{1-\frac{1}{\alpha}} t^{1-\alpha} \Gamma_\alpha^{\frac{1}{\alpha}}  \frac{1}{\alpha}(x^\alpha-v_\alpha t^\alpha)^{\frac{1}{\alpha}-1} \alpha v_\alpha t^{\alpha-1} D^\alpha_{x'} \Psi\\\nn
&+& \Gamma_\alpha^{1-\frac{1}{\alpha}} (t^\alpha-\frac{v_\alpha}{c^{2\alpha}} x^\alpha)^{1-\frac{1}{\alpha}} t^{1-\alpha} \Gamma_\alpha^{\frac{1}{\alpha}} \frac{1}{\alpha}(t^\alpha-\frac{v_\alpha}{c^{2\alpha}} x^\alpha)^{\frac{1}{\alpha}-1} \alpha t^{\alpha-1} D^\alpha_{t'} \Psi\\\label{first t}
&=&-v_\alpha \Gamma_\alpha  D^\alpha_{x'} \Psi-\Gamma_\alpha D^\alpha_{t'} \Psi.
\eea
Thus, we obtain 
\bea
\nn
D^\alpha_t D^\alpha_t \Psi &=&(-v_\alpha \Gamma_\alpha  D^\alpha_{x'}-\Gamma_\alpha D^\alpha_{t'})  (-v_\alpha \Gamma_\alpha  D^\alpha_{x'} \Psi-\Gamma_\alpha D^\alpha_{t'}\Psi), \\\label{2nd t}
&=& v_\alpha^2 \Gamma_\alpha^2  D^\alpha_{x'}  D^\alpha_{x'} \Psi - 2 v_\alpha \Gamma_\alpha^2   D^\alpha_{x'} D^\alpha_{t'} \Psi + \Gamma_\alpha^2 D^\alpha_{t'} D^\alpha_{t'}\Psi.
\eea
Substituting eqs.\eqref{2nd x},\eqref{2nd y'},\eqref{2nd z'}and \eqref{2nd t} in eq.\eqref{wave  eq 3d}, we have 
\bea
\nn
\Gamma_\alpha^2  D^\alpha_{x'}  D^\alpha_{x'} \Psi &-& 2 \Gamma_\alpha^2 \frac{v_\alpha}{c^{2\alpha}}  D^\alpha_{x'} D^\alpha_{t'} \Psi + \Gamma_\alpha^2 \frac{v_\alpha^2}{c^{4\alpha}} D^\alpha_{t'} D^\alpha_{t'}\Psi + D_{y'}^\alpha D_{y'}^\alpha + D_{z'}^\alpha D_{z'}^\alpha \\\nn 
&-& \frac{v_\alpha^2}{c^{2\alpha}} \Gamma_\alpha^2  D^\alpha_{x'}  D^\alpha_{x'} \Psi + 2 \frac{v_\alpha}{c^{2\alpha}}\Gamma_\alpha^2   D^\alpha_{x'} D^\alpha_{t'} \Psi - \frac{\Gamma_\alpha^2}{c^{2\alpha}} D^\alpha_{t'} D^\alpha_{t'}\Psi=0.
\eea
Rearranging,
\bea
\nn
\Gamma_\alpha^2 (1-\frac{v_\alpha^2}{c^{2\alpha}} )D^\alpha_{x'}  D^\alpha_{x'} \Psi + D_{y'}^\alpha D_{y'}^\alpha + D_{z'}^\alpha D_{z'}^\alpha  - \frac{\Gamma_\alpha^2}{c^{2\alpha}}  (1-\frac{v_\alpha^2}{c^{2\alpha}} ) D^\alpha_{t'} D^\alpha_{t'}\Psi =0.
\eea
Using $\Gamma_\alpha^2 (1-\frac{v_\alpha^2}{c^{2\alpha}} )=1$, we finally obtain
\bea
\nn
D^\alpha_{x'} D^\alpha_{x'} \Psi + D^\alpha_{y'} D^\alpha_{y'}  \Psi+ D^\alpha_{z'} D^\alpha_{z'} \Psi - \frac{1}{c^{2\alpha}} D^\alpha_{t'} D^\alpha_{t'} \Psi =0,
\eea
or
\bea
\nn
\nabla^{' 2\alpha} \Psi - \frac{1}{c^{2\alpha}} D^\alpha_{t'} D^\alpha_{t'} \Psi =0,
\eea
which shows that the $\alpha-$ wave equation is invariant under the $\alpha-$ Lorentz transformations. In the following three subsections, we provide three examples that are in support of the second postulate.
\subsection{Conformable Schrodinger equation}
The $\alpha-$ Lorentz transformation on conformable Schrodinger equation \cite{chung2020effect} is
\be
\label{TIME shro}
(\frac{\hat{p}_\alpha^2}{2m^\alpha}+V_\alpha(\hat{x}_\alpha))\Psi=i\hbar_\alpha^\alpha D^\alpha_t\Psi.
\ee
In $3+1$-dimensions, we have 
\be
\label{3d TIME shro }
-\frac{\hbar_\alpha^{2\alpha} }{2m^\alpha} [ D^\alpha_x D^\alpha_x + D^\alpha_y D^\alpha_y + D^\alpha_z D^\alpha_z] \Psi  +V_\alpha(\hat{x}_\alpha)\Psi=i\hbar_\alpha^\alpha D^\alpha_t\Psi.
\ee
Using $\alpha-$ Lorentz transformation by substituting from eqs.\eqref{2nd x},\eqref{2nd y'},\eqref{2nd z'}and \eqref{first t} in eq.\eqref{3d TIME shro }, we have 
\bea
\nn
&-&\frac{\hbar_\alpha^{2\alpha} }{2m^\alpha} [\Gamma_\alpha^2  D^\alpha_{x'}  D^\alpha_{x'} \Psi - 2 \Gamma_\alpha^2 \frac{v_\alpha}{c^{2\alpha}}  D^\alpha_{x'} D^\alpha_{t'} \Psi + \Gamma_\alpha^2 \frac{v_\alpha^2}{c^{4\alpha}} D^\alpha_{t'} D^\alpha_{t'}\Psi+ D_{y'}^\alpha D_{y'}^\alpha + D_{z'}^\alpha D_{z'}^\alpha \Psi]  \\\nn &+&V_\alpha(\hat{x}_\alpha)\Psi =i\hbar_\alpha^\alpha [-v_\alpha \Gamma_\alpha  D^\alpha_{x'} \Psi-\Gamma_\alpha D^\alpha_{t'} \Psi].
\eea
Thus, the conformable Schrodinger equation is not invariant under the $\alpha-$ Lorentz transformations.
\subsection{Conformable Gordon-Klein equation}
We firstly present the definition of conformable relativistic energy.\\
\textbf{Definition} The conformable relativistic energy is defined as 
\be
\label{energy}
E^{2\alpha}= p^{2\alpha}c^{2\alpha}+m^{2\alpha}c^{4\alpha}.
\ee
Quantization can be achieved by substituting for the conformable operators as 
$\hat{E}^\alpha=i\hbar_\alpha^\alpha D^\alpha_t$ and $\hat{p}^{\alpha}= -i\hbar_\alpha^\alpha \nabla^\alpha$ \cite{chung2020effect}.
The conformable Klein-Gordon equation is then
\be
\label{g&k}
\frac{1}{c^{2\alpha}} D_t^\alpha D_t^\alpha \Psi- \nabla^{2\alpha}\Psi+\frac{m^{2\alpha}c^{2\alpha}}{\hbar_\alpha^{2\alpha}} \Psi=0.
\ee
Substituting eqs.\eqref{2nd x},\eqref{2nd y'},\eqref{2nd z'}and \eqref{2nd t} in eq.\eqref{g&k}, we have 
\bea
\nn
&+&\frac{v_\alpha^2}{c^{2\alpha}} \Gamma_\alpha^2  D^\alpha_{x'}  D^\alpha_{x'} \Psi -2 \frac{v_\alpha}{c^{2\alpha}}\Gamma_\alpha^2   D^\alpha_{x'} D^\alpha_{t'} \Psi + \frac{\Gamma_\alpha^2}{c^{2\alpha}} D^\alpha_{t'} D^\alpha_{t'}\Psi+ \Gamma_\alpha^2  D^\alpha_{x'}  D^\alpha_{x'} \Psi \\\nn 
&+& 2 \Gamma_\alpha^2 \frac{v_\alpha}{c^{2\alpha}}  D^\alpha_{x'} D^\alpha_{t'} \Psi - \Gamma_\alpha^2 \frac{v_\alpha^2}{c^{4\alpha}} D^\alpha_{t'} D^\alpha_{t'}\Psi - D_{y'}^\alpha D_{y'}^\alpha - D_{z'}^\alpha D_{z'}^\alpha +\frac{m^{2\alpha}c^{2\alpha}}{\hbar_\alpha^{2\alpha}} \Psi=0.
\eea
Then,
\bea
\nn
\frac{\Gamma_\alpha^2}{c^{2\alpha}}  (1-\frac{v_\alpha^2}{c^{2\alpha}} ) D^\alpha_{t'} D^\alpha_{t'}\Psi -\Gamma_\alpha^2 (1-\frac{v_\alpha^2}{c^{2\alpha}} )D^\alpha_{x'}  D^\alpha_{x'} \Psi - D_{y'}^\alpha D_{y'}^\alpha - D_{z'}^\alpha D_{z'}^\alpha   +\frac{m^{2\alpha}c^{2\alpha}}{\hbar_\alpha^{2\alpha}} \Psi =0.
\eea
Thus, we have 
\be
\frac{1}{c^{2\alpha}} D^\alpha_{t'} D^\alpha_{t'} \Psi-D^\alpha_{x'} D^\alpha_{x'} \Psi - D^\alpha_{y'} D^\alpha_{y'}  \Psi -  D^\alpha_{z'} D^\alpha_{z'} \Psi   +\frac{m^{2\alpha}c^{2\alpha}}{\hbar_\alpha^{2\alpha}} \Psi =0,
\ee
or,
\be
\frac{1}{c^{2\alpha}} D^\alpha_{t'} D^\alpha_{t'} \Psi-\nabla^{' 2\alpha}\Psi -  D^\alpha_{z'} D^\alpha_{z'} \Psi   +\frac{m^{2\alpha}c^{2\alpha}}{\hbar_\alpha^{2\alpha}} \Psi =0.
\ee
Thus, the conformable Klein-Gordon equation is invariant under the $\alpha-$ Lorentz transformations.
\subsection{Four vector in conformable form}
We firstly present the definition of conformable position.\\
\textbf{Definition}. 1-The $\alpha-$ Covariant notation for position  $x_\mu^\alpha$ is defined as 
\be
\label{ Covariant notation}
x_\mu^\alpha = (x_0^\alpha,x_1^\alpha,x_2^\alpha,x_3^\alpha)= (c^\alpha t^\alpha,-x^\alpha,-y^\alpha,-z^\alpha).
\ee
2- The $\alpha-$ Contravariant notation for position  $x^{\mu,\alpha}$ is defined as 
\be
\label{ Contravariant notation}
x^{\mu,\alpha} = (x^{0,\alpha},x^{1,\alpha},x^{2,\alpha},x^{3,\alpha})= (c^\alpha t^\alpha,x^\alpha,y^\alpha,z^\alpha).
\ee
So, the relation between  $x_\mu^\alpha$ and $x^{\mu,\alpha}$ is given by 
\bea
x_\mu^\alpha = g_{\mu \nu} x^{\mu,\alpha} \quad or \quad  x^{\mu,\alpha}  = g^{\mu \nu} x_\mu^\alpha,
\eea
where $g_{\mu \nu}$ is a metric tensor \cite{pawar2018approach}, in Cartesian coordinates 
\bea
\nn
g_{\mu \nu}=g^{\mu \nu}=
\begin{pmatrix}
  1&0&0&0  \\
  0&-1&0&0  \\
 0&0&-1&0   \\
0&0&0&-1  
 \end{pmatrix}.
 \eea
Thus, the  displacement in conformable four vector is given by\\
 1- The $\alpha-$   Covariant displacement 
 \be
 \label{ Covariant displacement}
 d^\alpha x_\mu = ( d^\alpha x_0 , d^\alpha x_1 , d^\alpha x_2, d^\alpha x_3)= (c^\alpha  d^\alpha t, - d^\alpha x, - d^\alpha y, - d^\alpha z).
 \ee
 2- The $\alpha-$  Contravariant displacement
  \be
 \label{ alpha Covariant displacement}
 d^\alpha x^\mu = ( d^\alpha x^0 , d^\alpha x^1 , d^\alpha x^2, d^\alpha x^3)= (c^\alpha  d^\alpha t, d^\alpha x,  d^\alpha y, d^\alpha z).
 \ee
 So, the conformable differential line element is given by 
  \be
 \label{differential line element }
  d^\alpha x_\mu d^\alpha x^\mu = (c^{2\alpha}  d^{2\alpha} t, - d^{2\alpha} x, - d^{2\alpha} y, - d^{2\alpha} z).
 \ee 
Secondly, we present the definition of operators in conformable four vector.\\
 \textbf{Definition}. The dell operator in conformable four vector is defined as\\
 1- The $\alpha-$ Covariant dell operator is given by 
 \be
 \label{ Covariant dell operator}
 \partial_\mu^\alpha = ( \partial_0^\alpha, \partial_1^\alpha, \partial_2^\alpha, \partial_2^\alpha) = \frac{\partial^\alpha}{\partial (x^\mu)^\alpha} = (\frac{1}{c^\alpha}\frac{\partial^\alpha}{\partial t^\alpha},\nabla^\alpha).
 \ee
 2- The $\alpha-$ Contravariant dell operator is given by 
  \be
 \label{ Contravariant dell operator}
 \partial^{\mu,\alpha} = (\partial^{0,\alpha},\partial^{1,\alpha},\partial^{2,\alpha},\partial^{3,\alpha}) = \frac{\partial^\alpha}{\partial (x_\mu)^\alpha} = (\frac{1}{c^\alpha}\frac{\partial^\alpha}{\partial t^\alpha},-\nabla^\alpha).
 \ee
 Thus, the $\alpha-$ D’Alembert operator is given by 
 \be
 \label{ D’Alembert operator}
  \partial_\mu^\alpha \partial^{\mu,\alpha} =  \frac{1}{c^{2\alpha}}\frac{\partial^{2\alpha}}{\partial t^{2\alpha}} - \nabla^{2\alpha}.
 \ee
 So, through using $\alpha-$ D’Alembert operator the Conformable wave equation and the conformable Gorden-Klein equation are becomes  
 \be
 \partial_\mu^\alpha \partial^{\mu,\alpha}\Psi=0.
 \ee
  \be
 [\partial_\mu^\alpha \partial^{\mu,\alpha}+\frac{m^{2\alpha}c^{2\alpha}}{\hbar_\alpha^{2\alpha}}] \Psi=0.
 \ee
 Thus, we can obtained the energy-momentum four vector in conformable form as\\ 
 1- In $\alpha-$ Covariant form
 \be
 \label{ Covariant energy-momentum}
 P_\mu^\alpha = i \hbar_\alpha^\alpha \partial_\mu^\alpha = i \hbar_\alpha^\alpha  (\frac{1}{c^\alpha}\frac{\partial^\alpha}{\partial t^\alpha},\nabla^\alpha).
 \ee
  2- In $\alpha-$ Contravariant form
  \be
 \label{  Contravariant energy-momentum}
 P^{\mu,\alpha} = i \hbar_\alpha^\alpha  \partial^{\mu,\alpha} = i \hbar_\alpha^\alpha  (\frac{1}{c^\alpha}\frac{\partial^\alpha}{\partial t^\alpha},-\nabla^\alpha).
 \ee
In case  independent time of the conformable Schrodinger equation \cite{chung2020effect}, we get 
\be
i \hbar_\alpha^\alpha \frac{\partial^\alpha}{\partial t^\alpha} \Psi = E^\alpha \Psi.
\ee
So, the eqs (\ref{  Covariant energy-momentum}) and (\ref{ Contravariant energy-momentum}) are becomes
\be
 \label{ Covariant energy-momentum 2}
 P_\mu^\alpha = (\frac{E^\alpha}{c^\alpha},-\hat{p}_\alpha).
 \ee
 and
  \be
 \label{  Contravariant energy-momentum 2}
 P^{\mu,\alpha} =  (\frac{E^\alpha}{c^\alpha},\hat{p}_\alpha).
 \ee
 Where $\hat{p}_\alpha$ is called $\alpha-$ momentum operator \cite{chung2020effect}, in one dimension is equal $\hat{p}_\alpha = -i \hbar_\alpha^\alpha D^\alpha_x $ and in 3-D it is equal  $\hat{p}_\alpha = -i \hbar_\alpha^\alpha \nabla^\alpha $. 
 \subsection{The $\alpha$-Lorentz transformation in Minkowski Space}
 The $\alpha$-Lorentz transformation in Minkowski Space is given by\\ 
 1- In the $\alpha-$ Contravariant form 
 \be
 \label{LT MIN CONTRA}
 x^{'\mu,\alpha} = ~^\alpha\!\Lambda^\mu_\nu x^{\nu,\alpha}
 \ee
where $~^\alpha\!\Lambda^\mu_\nu$ is $\alpha$-tensor and defined as
\bea
\nn
~^\alpha\!\Lambda^\mu_\nu&=&\frac{\partial^\alpha }{\partial (x^{\nu })^\alpha} (\frac{x^{'\mu,\alpha}}{\alpha})=
 \begin{pmatrix}
  \Gamma_\alpha & - \Gamma_\alpha \beta^\alpha &0&0  \\
 - \Gamma_\alpha \beta^\alpha & \Gamma_\alpha &0&0  \\
 0&0&1&0   \\
0&0&0&1  
 \end{pmatrix},
\eea
where $\beta^\alpha= \frac{v^\alpha}{c^\alpha}$ and the inverse of it is $x^\nu =  (~^\alpha\!\Lambda^\mu_\nu)^{-1} x^{'\mu,\alpha}$.\\
2-The $\alpha-$ Covariant form
\be
\label{LT MIN CO}
x_\mu^{' \alpha} =  ~^\alpha\!\Lambda_\mu^\nu x_\nu^\alpha,
\ee
where $~^\alpha\!\Lambda_\mu^\nu$ is given by
\bea
\nn
~^\alpha\!\Lambda_\mu^\nu =\frac{\partial^\alpha }{\partial (x^{\mu })^\alpha} (\frac{x^{'\nu,\alpha}}{\alpha})= \begin{pmatrix}
  \Gamma_\alpha &  \Gamma_\alpha \beta^\alpha &0&0  \\
  \Gamma_\alpha \beta^\alpha & \Gamma_\alpha &0&0  \\
 0&0&1&0   \\
0&0&0&1  
 \end{pmatrix}
\eea
\textbf{Proof.} through using this equation
\bea
\label{main}
x_\mu^{'\alpha} = g_{\mu s} x^{'s,\alpha},
\eea
can write $x^{'s,\alpha}$ using The $\alpha$-Lorentz transformation in Contravariant form as eq.(\ref{LT MIN CONTRA}) 
\bea
x^{'s,\alpha} =  ~^\alpha\!\Lambda_\theta^s x^{\theta,\alpha}.
\eea
Substituting it  in eq.(\ref{main}), we get 
\bea
\label{main 1}
x_\mu^{'\alpha} = g_{\mu s} ~^\alpha\!\Lambda_\theta^s x^{\theta,\alpha}.
\eea
So, can write $x^\theta$ as 
\bea
x^{\theta,\alpha} = g^{\theta \nu} x_{\nu}^\alpha.
\eea
Substituting it  in eq.(\ref{main 1}), we get 
\bea
\label{main2}
x_\mu^{'\alpha} = g_{\mu s} ~^\alpha\!\Lambda_\theta^s g^{\theta \nu} x_{\nu}^\alpha.
\eea
Thus, $g_{\mu s} ~^\alpha\!\Lambda_\theta^s g^{\theta \nu}$ are three matrices
\bea
\nn
g_{\mu s} ~^\alpha\!\Lambda_\theta^s g^{\theta \nu} &=& \begin{pmatrix}
  1&0&0&0  \\
  0&-1&0&0  \\
 0&0&-1&0   \\
0&0&0&-1  
 \end{pmatrix} 
  \begin{pmatrix}
  \Gamma_\alpha & - \Gamma_\alpha \beta^\alpha &0&0  \\
 - \Gamma_\alpha \beta^\alpha & \Gamma_\alpha &0&0  \\
 0&0&1&0   \\
0&0&0&1  
 \end{pmatrix}
 \begin{pmatrix}
  1&0&0&0  \\
  0&-1&0&0  \\
 0&0&-1&0   \\
0&0&0&-1  
 \end{pmatrix}\\\nn &=& \begin{pmatrix}
  \Gamma_\alpha &  \Gamma_\alpha \beta^\alpha &0&0  \\
  \Gamma_\alpha \beta^\alpha & \Gamma_\alpha &0&0  \\
 0&0&1&0   \\
0&0&0&1  
 \end{pmatrix} = ~^\alpha\!\Lambda_\mu^\nu
\eea
Taking the inverse of $~^\alpha\!\Lambda^\mu_\nu$, we get 
\bea
(~^\alpha\!\Lambda^\mu_\nu)^{-1} = \begin{pmatrix}
  \Gamma_\alpha &  \Gamma_\alpha \beta^\alpha &0&0  \\
  \Gamma_\alpha \beta^\alpha & \Gamma_\alpha &0&0  \\
 0&0&1&0   \\
0&0&0&1  
 \end{pmatrix}  = g_{\mu s} ~^\alpha\!\Lambda_\theta^s g^{\theta \nu} = ~^\alpha\!\Lambda_\mu^\nu
\eea
So, the eq.(\ref{main2}) is become the eq.(\ref{LT MIN CO}).\\
\subsection{Conformable Dirac Equation}
In Mozaffari et.al \cite{mozaffari2018investigation}, the Dirac equation by using the conformable derivative is investigated and it is given by
\be
\label{Conformable Dirac Equation}
[i\gamma^\mu \partial_\mu^\alpha - m^\alpha]\Psi(x^{\mu,\alpha})=0,
\ee
where $\gamma^\mu$ are the famous $\gamma$ matrices of Dirac equation \cite{bjorken1964relativistic}.
Similarly, the Dirac equation is Lorentz covariant
\be
\label{Conformable Dirac Equation in s}
[i\gamma^\nu \partial_\nu^{\prime\alpha}- m^\alpha]\Psi^\prime (x^{\prime\nu,\alpha}) =0,
\ee
but when we do a Lorentz transformation, the wave function changes. So, We require that the transformation between $\Psi$ and $\Psi^\prime$ be linear since the Dirac equation and Lorentz transformation are linear.
\bea
\label{psi}
\Psi^\prime (x^{\prime\alpha}) = S \Psi(x^\alpha),
\eea
 where S denotes an x-independent matrix whose properties must be found. The Dirac equation in Lorentz covariance indicates that the $\gamma$ matrices are identical in both frames. Using 
 \bea
 \label{par}
 \partial_\mu^{\alpha} = ~^\alpha\!\Lambda_\mu^\nu \partial_\nu^{\prime\alpha}.
\eea
From  eq.(\ref{psi}) we found $\Psi(x^\alpha)= S^{-1} \Psi^\prime (x^{\prime\alpha}) $ and substituting in eq.(\ref{Conformable Dirac Equation}), we get 
\be
\label{Conformable Dirac Equation1 1}
[i\gamma^\mu \partial_\mu^{\alpha} - m^\alpha]S^{-1} \Psi^\prime (x^{\prime\alpha})=0.
\ee
After substituting eq.(\ref{par}) in it. Multiply it  with $S$ from the left, we get 
\be
\label{Conformable Dirac Equation1 2}
[i S\gamma^\nu S^{-1} ~^\alpha\!\Lambda_\mu^\nu \partial_\nu^{\prime\alpha}- m^\alpha] \Psi^\prime (x^{\prime\alpha})=0.
\ee
Comparing eq.(\ref{Conformable Dirac Equation1 2}) with eq.(\ref{Conformable Dirac Equation in s}), we obtained 
\be
\label{relation}
S\gamma^\mu S^{-1} ~^\alpha\!\Lambda_\mu^\nu = \gamma^\nu
\ee
So, $\gamma^\mu  ~^\alpha\!\Lambda_\mu^\nu = S^{-1} \gamma^\nu S$. The inverse Lorentz transformation must, correspond to the inverse S \cite{nikolic2014not}, we get 
\be
\label{relation1}
\gamma^\mu  (~^\alpha\!\Lambda_\mu^\nu)^{-1} = S \gamma^\nu S^{-1}
\ee
So we demonstrated that the conformable Dirac equation is covariant in $\alpha$-Lorentz transformation, and for more information on the S matrix, see this ref \cite{bjorken1964relativistic,nikolic2014not}.
\section{Summary and conclusions}
In this work, we have investigated the deformation of Einstein's special relativity using the concept of conformable derivative. Within this context, the two postulates of the theory were re-stated. Then, the conformable addition of velocity laws were derived and used to verify the constancy of the speed of light.  
The invariance principle of the laws of physics was demonstrated for some typical illustrative examples, namely, the conformable wave equation, the conformable Schrodinger equation, and the conformable Gordon-Klein equation.
\bibliography{ref-v3}
\bibliographystyle{IEEEtran}

\end{document}